\documentclass[letterpaper]{jpsj3}
\usepackage{txfonts}
\usepackage{bm}
\usepackage{xcolor}

\usepackage[whole]{bxcjkjatype}

\title{Subgradient Method using Quantum Annealing for Inequality-Constrained Binary Optimization Problems}

\author{Taisei Takabayashi$^{1 *}$, Takeru Goto$^{1, 2}$ and Masayuki Ohzeki$^{1, 3, 4}$}
\inst{Graduate School of Information Sciences, Tohoku University, Miyagi 980-8564, Japan$^1$ \\
Innovative Research Excellence, Honda R \& D Co., Ltd., Tokyo 107-6238, Japan$^2$ \\
Department of Physics, Institute of Science Tokyo, 152-8551, Japan$^3$ \\
Sigma-i Co., Ltd., Tokyo, 108-0075, Japan$^4$
} 

\abst{
Quantum annealing is a generic solver for combinatorial optimization problems that utilizes quantum fluctuations. 
Recently, there has been extensive research applying quantum annealers, which are hardware implementations of quantum annealing. 
Since quantum annealers can only handle quadratic unconstrained binary optimization problems, to solve constrained combinatorial optimization problems using quantum annealers, the constraints must be incorporated into the objective function. 
One such technique is the Ohzeki method, which employs a Hubbard-Stratonovich transformation to relax equality constraints, and its effectiveness for large-scale problems has been demonstrated numerically.
This study applies the Ohzeki method to combinatorial optimization problems with inequality constraints. 
We show that inequality constraints can be relaxed into a similar objective function through statistical mechanics calculations similar to those for equality constraints. 
In addition, we evaluate the performance of this method in a typical inequality-constrained combinatorial optimization problem, the quadratic knapsack problem.
}

\begin{document}
\maketitle

\section{Introduction.}
Quantum annealing (QA) is a general-purpose method for solving combinatorial optimization problems using quantum fluctuations\cite{kadowaki1998quantum}. 
Recently, with the development of the quantum annealer by D-Wave Systems, which implements QA in a commercial device, Quantum annealing has attracted attention as a new computational technique. Research on applying quantum annealers, including D-Wave machines, has been conducted across various industries. For instance, QA has been proposed for use in traffic flow optimization\cite{neukart2017traffic, inoue2021traffic, shikanai2023trafficsignaloptimizationusing}, finance\cite{rosenberg2016solving, venturelli2019reverse}, maze generation\cite{Ishikawa_2023}, manufacturing\cite{ohzeki2019control, Haba2022}, preprocessing in material experiments\cite{Tanaka2023, Doi_2023}, steel manufacturing\cite{yonaga_quantum_2022}, decoding problems\cite{ide2020maximum, Arai2021code}, and algorithms in machine learning\cite{amin2018quantum, o2018nonnegative, sato_assessment_2021, Urushibata2022, hasegawa2023, goto2023onlinecalibrationschemetraining}. In addition, the benchmark test of model-based
Bayesian optimization using QA was performed\cite{Koshikawa2021}. A comparative study of QA with other optimization solvers has been conducted\cite{Oshiyama2022}. Moreover, there are discussions about the quantum effects when multiple optimal solutions exist\cite{Yamamoto2020, Maruyama2021}. 

Quantum annealers are primarily designed to solve quadratic unconstrained binary optimization (QUBO) problems\cite{glover_quantum_2022}. Therefore, when solving optimization problems with equality or inequality constraints, it is necessary to relax these constraints into the objective function. The common method for this is the penalty method. In the penalty method, equality constraints are relaxed into the objective function as squared error penalty terms. However, these penalty terms generate quadratic terms, leading to dense coupling between variables.
Given that the physical qubits in devices like the D-Wave machine are sparsely connected, this dense coupling requires multiple qubits to represent a single variable, significantly reducing the size of problems that can be represented and solved on such devices. Moreover, on the D-Wave machine, a group of qubits representing a single variable, known as a chain, may sometimes exhibit different values. This phenomenon, called chain break, is one factor that reduces the precision of the solutions obtained from the D-Wave machine. The dense coupling caused by the penalty method tends to increase the occurrence of chain breaks, thereby affecting the optimization performance.

To overcome this obstacle, a method known as the Ohzeki method has been proposed, which converts the quadratic penalty terms into linear terms\cite{Ohzeki2020}. This method defines a partition function on the Gibbs-Boltzmann distribution for the objective function that includes the constraints and then applies the Hubbard-Stratonovich transformation\cite{Hubbard1959, Stratonovich1957} to convert the quadratic penalty terms into linear terms. This transformation allows larger problems to be represented on the hardware of quantum annealers. Moreover, the saddle-point method calculation for the partition function yields update rules for the newly introduced auxiliary variables. These auxiliary variables correspond to Lagrange multipliers in mathematical optimization, making the Ohzeki method a way to sample solutions to the Lagrangian relaxation problem\cite{geoffrion_lagrangean_1974} using quantum annealers.

The original research assumed equality constraints. However, many combinatorial optimization problems have inequality constraints. This study demonstrates that the Ohzeki method can also be applied to optimization problems with inequality constraints. By applying a transformation similar to the Hubbard-Stratonovich transformation to the partition function defined for an inequality-constrained optimization problem, we derive an objective function similar to that for equality constraints. 

To evaluate the performance of the derived algorithm, we use the quadratic knapsack problem (QKP), a representative example of an inequality-constrained combinatorial optimization problem. For the sampling needed in the expectation calculation under the Boltzmann distribution, we employed the Markov-chain Monte-Carlo (MCMC) method and simulated quantum annealing (SQA), a method that classically simulates QA using quantum Monte Carlo techniques\cite{suzuki_monte_1977}. The properties of SQA as a Boltzmann sampler have been investigated in the literature \cite{PhysRevA.104.022607}. In addition, we prepared two methods for comparison with our algorithm. The first is a naive subgradient method, which exactly solves the Lagrangian relaxation problem, unlike sampling using MCMC or SQA. The second is a greedy method for the QKP\cite{BILLIONNET1996310}. 

We conducted experiments while varying the number of decision variables and the density of interactions between the variables in the QKP. 
The results showed that although the Ohzeki method and the naive subgradient method were less accurate than the greedy method except for certain cases, the Ohzeki method generally provided better solutions than the naive one, especially by using SQA. 

This result demonstrates the effectiveness of sampling in methods that combine Lagrangian relaxation problems with subgradient methods.

The remainder of this paper is structured as follows: Section 2 introduces QA and the Ohzeki method for optimization problems with equality constraints as background. In Sect. 3, we extend the Ohzeki method to handle inequality constraints. Sect. 4 presents numerical experiments using the QKP to verify the proposed method's performance. Finally, in Sect. 5, we discuss the results and conclude the paper.

\section{Background}
\subsection{Quantum Annealing and QUBO}
Quantum annealing (QA) leverages quantum mechanical phenomena to search for the ground state of the following Ising model:
\begin{equation}
    \mathcal{H}_{\rm Ising}({\bm s})=\frac{1}{2} \sum_{i\ne j}^{N}J_{ij}s_i s_j +\sum_{i=1}^{N}h_i s_i, \quad {\bm s} \in \{-1,+1\}^N,
    \label{Ising model}
\end{equation}
where $s_i \in \{+1, -1\}$ represents the $i$-th spin, $N$ is the total number of spins, $J_{ij}$ denotes the interaction between adjacent spins $s_i$ and $s_j$, and $h_i$ represents the local longitudinal magnetic field acting on the $i$-th spin. 
The ground state of the Ising model corresponds to the spin configuration that minimizes this Hamiltonian. 
Since many combinatorial optimization problems can be represented as this Ising model, QA solves these problems by finding the ground state of the corresponding Ising Hamiltonian.

It is common to express combinatorial optimization problems as the following quadratic unconstrained binary optimization (QUBO) problem:
\begin{equation}
\min_{{\bm x}\in \{0,1\}^N} \quad \sum_{i=1}^{N}\sum_{j=1}^{N} Q_{ij} x_i x_j, 
\label{qubo}
\end{equation}
where $x_i$ is a binary variable, and $Q$ is an $N\times N$ symmetric matrix representing the QUBO matrix. 
Here QUBO can be transformed into Ising models by the variable transformation $s_i = 2 x_i - 1$, allowing QA to handle QUBO problems.

Although QUBO does not inherently include constraints, many combinatorial optimization problems involve constraints. 
The constrained optimization problem can be defined as follows:
\begin{equation}
\begin{aligned}
\min_{{\bm x}\in \{0,1\}^N} \quad & f_0({\bm x}) \\
\textrm{s.t.} \quad & F_k({\bm x}) = C_k, \quad (k=1, \dots, K), 
\label{problem under equality constraint}
\end{aligned}
\end{equation}
where $F_k({\bm x})$ is a function of the binary variables ${\bm x}=(x_1, x_2, \dots, x_N)$ and $F_k({\bm x}) = C_k$ represents equality constraints. $K$ is the number of constraints. To express this problem as a QUBO, it is necessary to include the equality constraints in the objective function, typically using a penalty method. The penalty method expresses the equality constraints as squared penalty terms:
\begin{equation}
\min_{{\bm x}\in \{0,1\}^N} \quad f_0({\bm x}) + \frac{1}{2} \sum_{k=1}^{K} \lambda_k \left( 
F_k ({\bm x}) - C_k \right)^2,
\label{penalty method for equality constraints}
\end{equation}
where $\bm \lambda$ is the penalty coefficient that controls the strength of the penalty.

However, this penalty method has a problem: introducing quadratic terms leads to dense coupling between variables. 
For example, if the constraint term $F_k({\bm x})$ is a linear combination of ${\bm x}$, i.e., $F_k({\bm x})=\sum_{i=1}^N a_{k i} x_i$ (where ${\bm a_k}$ is the coefficient vector for constraint $k$), then expanding the penalty term in Eq. \eqref{penalty method for equality constraints} yields a quadratic term $\sum_{i=1}^N \sum_{j=1}^N a_{k i} a_{k j} x_i x_j$. 
This implies that each logical variable ${\bm x}$ is coupled with every other variable, meaning that the QUBO matrix $Q$ will consist of non-zero elements only. 
This is problematic for quantum annealers like the D-Wave machine, with sparse qubit connections. 
When the logical variables are densely coupled, representing them on the quantum annealer requires more qubits, thus limiting the size of the problem that can be represented and solved.

To overcome this obstacle, the Ohzeki method has been proposed, reformulating the quadratic penalty terms into linear terms, thereby reducing the density of couplings and enabling the quantum annealer to handle larger problems more effectively. 
In the next chapter, we will briefly introduce the Ohzeki method for the optimization problem with equality constraints \eqref{problem under equality constraint}.

\subsection{Ohzeki method for equality-constrained problems}

The Ohzeki method transforms the quadratic penalty terms in the QUBO \eqref{penalty method for equality constraints} into linear terms. We apply the Hubbard-Stratonovich transformation\cite{Hubbard1959, Stratonovich1957} to the partition function defined for Eq. \eqref{penalty method for equality constraints}. 
As a result, the new objective function is expressed as follows:
\begin{equation}
\begin{aligned}
H({\bm x};{\bm \nu}) = f_0({\bm x})- \sum_{k=1}^{K}\nu_k F_k({\bm x}),
\label{new objective function for equality constraints}
\end{aligned}
\end{equation}
where ${\bm \nu}$ represents the auxiliary variables introduced due to the Hubbard-Stratonovich transformation. 
In this new objective function, the terms related to the constraints are linear and, therefore, do not result in fully connected couplings between variables. 
This effectively resolves the issue of reduced problem size due to the sparse graph structure of the quantum processing unit (QPU). 
The solutions sampled from the quantum annealer follow the Boltzmann distribution given by:
\begin{equation}
\begin{aligned}
Q({\bm \nu}) = \frac{1}{Z({\bm \nu})} \exp{(-\beta H({\bm x};{\bm \nu}))},
\label{effective boltzmann distribution for equality constraints}
\end{aligned}
\end{equation}
with the partition function defined as:
\begin{equation}
\begin{aligned}
Z({\bm \nu})=\sum_{{\bm x}} \exp{(-\beta H({\bm x};{\bm \nu}))},
\label{effective partition function for equality constraints}
\end{aligned}
\end{equation}
where $\beta$ is the inverse temperature parameter. The auxiliary variables ${\bm \nu}$ must be adjusted to satisfy the equality constraints. We apply the steepest ascent method for updating $\bm \nu$. The update rule for ${\bm \nu}$ is derived using the saddle-point method in the limit $\beta \rightarrow + \infty$, and is expressed as follows:
\begin{equation}
\begin{aligned}
\nu_k^{(t+1)} \leftarrow \nu_k^{(t)}+\eta^{(t)}(C_k - \langle F_k({\bm x}) \rangle_{Q({\bm \nu^{(t)}})}) \quad (t=1, 2, \dots), 
\label{update rule for equality constraint}
\end{aligned}
\end{equation}
where $\langle F_k({\bm x}) \rangle_{P({\bm x^{(t)}})}$ denotes the expected value of $F_k({\bm x})$ under the distribution $Q({\bm \nu^{(t)}})$. The index $t$ represents the iteration step, and $\eta$ is a positive parameter. Calculating the expected value $\langle F_k({\bm x}) \rangle_{P({\bm x^{(t)}})}$ is generally difficult because it requires the calculation of the partition function $Z({\bm \nu})$. However, since quantum annealers can sample from the Boltzmann distribution \eqref{effective boltzmann distribution for equality constraints}, this expected value is equal to the average $\frac{1}{S} \sum_{s=1}^{S}F_k({\bm x_s^{(t)}})$, where $({\bm x}_1^{(t)}, {\bm x}_2^{(t)}, \cdots, {\bm x}_S^{(t)})$ are the solutions sampled from the quantum annealer at step $t$, and $S$ is the number of samples. 

The Ohzeki method introduced above assumes that the constraints are in the form of equalities, as expressed in Eq. \eqref{problem under equality constraint}. 
However, many combinatorial optimization problems involve inequality constraints. 
The following chapter will extend the Ohzeki method to handle optimization problems with inequality constraints. 
We will define a partition function for such problems and show that a similar algorithm can be derived using the saddle-point method.


\section{Method}

This section demonstrates that an update rule similar to Eq. \eqref{update rule for equality constraint} can also be derived for optimization problems with inequality constraints. An inequality-constrained optimization problem can be defined as follows:
\begin{equation}
\begin{aligned}
\min_{{\bm x}\in \{0,1\}^N} \quad & f_0({\bm x}) \\
\textrm{s.t.} \quad & F_k({\bm x}) \le C_k, \quad (k=1, 2, \dots, K).
\label{problem under inequality constrained problem}
\end{aligned}
\end{equation}
Then, the partition function corresponding to this optimization problem can be expressed as:
\begin{equation}
\begin{aligned}
Z = \sum_{{\bm x}} \exp{\left( -\beta f_0({\bm x}) \right)} \prod_{k=1}^{K}\Theta \left(C_k - F_k({\bm x})\right),
\end{aligned}
\end{equation}
where $\Theta (a)$ is the Heaviside step function, which equals 1 if $a \ge 0$ and $0$ if $a<0$. 
Therefore, the partition  function $Z$ is defined only for configurations ${\bm x}$ that satisfy the inequality constraints $F_k({\bm x}) \le C_k(k=1,\cdots, K)$. The Heaviside step function can be represented as an integral:
\begin{equation}
\begin{aligned}
\Theta (y-\kappa) = \int_{\kappa}^{+\infty}\frac{d \xi}{2\pi}\int_{-\infty}^{+\infty}d z \exp{\left( - i z(\xi - y)\right)},
\end{aligned}
\end{equation}
which allows the partition function $Z$ to be rewritten as:
\begin{equation}
\begin{aligned}
Z = \sum_{{\bm x}} \exp{\left( -\beta f_0({\bm x}) \right)} \prod_{k=1}^{K} \int_{0}^{+\infty} \frac{d \xi_k}{2\pi}\int_{-\infty}^{+\infty} d z_k \exp{\left( - i \sum_{k=1}^{K} z_k (\xi_k -(C_k - F_k({\bm x}))) \right)}.
\label{partition function after step-function transformation}
\end{aligned}
\end{equation}
By performing the variable transformation $z_k \rightarrow -i \beta \mu_k$, the resulting partition function can be expressed as:
\begin{equation}
\begin{aligned}
Z \propto \sum_{{\bm x}} \prod_{k=1}^{K} \int_{0}^{+\infty} d \xi_k \int_{-\infty}^{+\infty} d\mu_k \exp{\left( -\beta \left(f_0({\bm x}) + \sum_{k=1}^{K} \mu_k (\xi_k + F_k({\bm x}) - C_k) \right)\right)}.
\label{partition function after variable transformation}
\end{aligned}
\end{equation}
Thus, we consider minimizing the effective Hamiltonian defined as:
\begin{equation}
\begin{aligned}
H({\bm x}; {\bm \mu}, {\bm \xi}) = f_0({\bm x}) + \sum_{k=1}^{K} \mu_k (\xi_k + F_k({\bm x}) - C_k).
\end{aligned}
\end{equation}
Continuing from the effective Hamiltonian \eqref{partition function after variable transformation}, the partition function can be further rewritten as:
\begin{equation}
\begin{aligned}
Z = \prod_{k=1}^{K} \int_{0}^{+\infty} d \xi_k \int_{-\infty}^{+\infty} d\mu_k \exp{\left( -\beta \left( \sum_{k=1}^{K} \mu_k (\xi_k - C_k) - \frac{1}{\beta} \log{Z({\bm \mu})} \right) \right)},
\end{aligned}
\end{equation}
where $Z({\bm \mu})$ is the effective partition function defined as:
\begin{equation}
\begin{aligned}
Z({\bm \mu}) = \sum_{{\bm x}} \exp{\left( -\beta \left( f_0({\bm x}) + \sum_{k=1}^{K} \mu_k F_k({\bm x}) \right) \right)}.
\end{aligned}
\end{equation}
Therefore, the effective Hamiltonian for ${\bm \mu}$ and ${\bm \xi}$ is obtained as:
\begin{equation}
\begin{aligned}
H ({\bm \mu}, {\bm \xi}) = \sum_{k=1}^{K} \mu_k (\xi_k - C_k) - \frac{1}{\beta} \log{Z({\bm \mu})}.
\label{effective hamiltonian for inequality constraints}
\end{aligned}
\end{equation}
Next, we seek the saddle point of this effective Hamiltonian in the limit $\beta \rightarrow + \infty$, which corresponds to the minimizer of the effective Hamiltonian:
\begin{equation}
\begin{aligned}
\frac{\partial}{\partial \xi_k} H ({\mu_k}, {\xi_k}) = \mu_k = 0,
\label{saddle point inequality1}
\end{aligned}
\end{equation}
\begin{equation}
\begin{aligned}
\frac{\partial}{\partial \mu_k} H ({\mu_k}, {\xi_k}) = \xi_k - C_k + \langle F_k({\bm x})	\rangle_{Q({\bm \mu})} = 0,
\label{saddle point inequality2}
\end{aligned}
\end{equation}
where the brackets $\langle \cdot \rangle$ represent the expectation value under the probability distribution $Q({\bm \mu})$.
\begin{equation}
\begin{aligned}
Q({\bm \mu}) = \frac{1}{Z ({\bm \mu})}   \exp{\left( -\beta \left( f_0({\bm x}) + \sum_{k=1}^{K} \mu_k F_k({\bm x}) \right) \right)}.
\end{aligned}
\end{equation}
Thus, the new objective function that requires sampling to compute the expectation values is given by:
\begin{equation}
\begin{aligned}
H({\bm x};{\bm \mu}) = f_0({\bm x}) + \sum_{k=1}^{K}\mu_k F_k({\bm x}) \quad ({\bm \mu}\ge {\bm 0}),
\label{new objective function for inequality constraints}
\end{aligned}
\end{equation}
which is equivalent to the objective function in the equality constraint case \eqref{new objective function for equality constraints} except for the range of the artificial variables $\bm \mu$. 
To reach the saddle point, we apply the steepest ascent method. 
The update rule is based on Eqs. \eqref{saddle point inequality1} and \eqref{saddle point inequality2} can be described as:
\begin{equation}
\begin{aligned}
\mu_k^{(t+1)} \leftarrow \max \left\{0, \ \mu_k^{(t)} + \eta^{(t)} \left(\left\langle F_k({\bm x})	\right\rangle_{Q({\bm  \mu^{(t)}})} - C_k \right)\right\}\quad (t=1, 2, \dots),
\label{sub-gradient update rule}
\end{aligned}
\end{equation}
where $\eta^{(t)}$ is the step size. The derivation of this rule is described in Appendix A. The above led to an updated rule of the Ohzeki method for the inequality-constrained optimization problem.

This method has two advantages. 
The first advantage is that it does not require the introduction of additional variables to represent the inequality constraints. Generally, when expressing inequality constraints in a QUBO problem, auxiliary variables $z_k\in[0, C_k]$ are introduced as $F_k({\bm x}) + z_k = C_k$, and the problem is converted into one with equality constraints, which can then be handled by the penalty method. 
However, because the auxiliary variables $z_k$ are not binary, they must be expanded into a sum of multiple binary variables to be represented in the QUBO. 
This expansion increases the number of binary variables, which requires additional qubits. This reduces the precision of the solutions obtained from the quantum annealer and limits the problem size that can be handled. 
In contrast, the Ohzeki method does not require such auxiliary variables, allowing for more efficient use of qubits. As with equality constraints, the Ohzeki method relaxes the constraint term into a linear form, thereby reducing the number of quadratic terms in the QUBO formulation compared to the slack variable approach. This reduction facilitates embedding into sparsely connected devices such as quantum annealers. Appendix B provides supplementary experimental results comparing the number of quadratic terms in the QUBO for both methods.

The second advantage is that the constraints are relaxed as linear terms. Therefore, the method can be applied even when the constraint $F_k({\bm x})$ includes quadratic terms, such as $F_k({\bm x}) = \sum_{i=1}^{N} \sum_{j=1}^{N} A_{k i j} x_i x_j$. 

The next chapter will validate the proposed algorithm using the quadratic knapsack problem (QKP), a typical inequality-constrained combinatorial optimization problem. Specifically, we will investigate the dependence of the algorithm's performance on the problem size and the density of the objective function $f_0 ({\bm x})$.

\section{Experiments}
\subsection{Parameter settings of the algorithm}
In this section, we evaluate the algorithm's performance introduced in Sect. 3. First, we describe the parameter settings of the algorithm. The step size $\eta^{(t)}$ in Eq. \eqref{sub-gradient update rule} is determined as follows\cite{fisher_lagrangian_1981}:
\begin{equation}
\begin{aligned}
\eta^{(t)} = \tau \frac{f_0^{\rm UB}({\bm x}) - \left\{\left\langle f_0({\bm x}) \right\rangle_{Q({\bm  \mu^{(t)}})} + \sum_{k=1}^{K} \left(\left\langle F_k({\bm x}) \right\rangle_{Q({\bm  \mu^{(t)}})} - C_k\right)\right\}} {\sum_{k=1}^{K} \left(\langle F_k({\bm x}) \rangle_{Q({\bm  \mu^{(t)}})} - C_k\right)^2},
\label{sub-gradient eta update rule}
\end{aligned}
\end{equation}
where $f_0^{\rm UB}({\bm x})$ is an upper bound on the objective function obtained by some heuristic method. 
In this experiment, we use the greedy method proposed for the QKP\cite{BILLIONNET1996310} to calculate $f_0^{\rm UB}({\bm x})$. 
We start with $\tau\ge 0$ and set $\tau=0.5$. If the best feasible solution ${\bm x}_{\rm feas}^{(t)}$ sampled from the quantum annealer does not improve for 10 steps, we reduce $\tau$ by half, i.e., $\tau \leftarrow 0.5 \times \tau$. Here, ${\bm x}_{\rm feas}^{(t)}$ represents the best feasible solution sampled at step $t$ from the quantum annealer. 
If $\tau$ is updated and there is no improvement in ${\bm x}_{\rm feas}^{(t)}$ for another 10 steps, we further reduce $\tau$ by half\cite{Shunji_Umetani2007KJ00004805415}.

The stopping conditions for the subgradient method are set as follows. 
The algorithm stops when either of the following conditions is met at step $t$:
\begin{enumerate}
  \item The number of steps $t$ exceeds $t_{\rm max}$.
  \item $\tau < \tau_{\rm min}$.
  \item $\sqrt{ \sum_{k=1}^{K}\left(\langle F_k({\bm x}) \rangle_{Q({\bm  \mu^{(t)}})} - C_k \right)^2 } < \epsilon$.
\end{enumerate}

Here, $t_{\rm max}$, $\tau_{\rm min}$, $\epsilon$ are predetermined parameters. We set $t_{\rm max}=50$, $\tau_{\rm min}=0.01$, $\epsilon=0.001$.

\subsection{Problem setting}
We use the quadratic knapsack problem (QKP), a typical example of an inequality-constrained combinatorial optimization problem for the experiment. The QKP is formulated as follows:
\begin{equation}
\begin{aligned}
\max_{{\bm x}\in \{0,1\}^N} \quad & \sum_{i=1}^N \sum_{j=1}^N P_{i j} x_i x_j \\
\textrm{s.t.} \quad & \sum_{i=1}^N w_i x_i \le c, 
\label{QKP}
\end{aligned}
\end{equation}
where $P=\{p_{i,j}\}\in \mathbb{Z}_{+}^{N\times N}$ is the symmetry profit matrix, ${\bm w}=\{w_i\}\in \mathbb{Z}_{+}^{N}$ is the weight vector, and $c\in \mathbb{Z}$ is the capacity. Following the literature\cite{gallo_quadratic_1980}, we generate these parameters randomly. The diagonal elements $P_{ii}$ of the profit matrix are all non-zero, while the off-diagonal elements are non-zero with probability $\Delta$. 
The values of the non-zero elements are drawn from a uniform distribution in the range $[1, 100]$. 
Thus, the parameter $\Delta$ represents the density, and when $\Delta$ is close to $1$, the objective function is dense. 
The elements of the weight vector ${\bm w}$ are chosen randomly from the range $[1, 50]$, and the capacity $c$ is randomly selected from the range $[50, \sum_{i=1}^{N}w_i]$.

The QKP has a single constraint, i.e., $K=1$. 
Compared to Eq. \eqref{problem under inequality constrained problem}, we have $f_0({\bm x})=-\sum_{i=1}^N \sum_{j=1}^N P_{i j} x_i x_j$, $F_1({\bm x})=\sum_{i=1}^N w_i x_i$, $C_1=c$. 
We prepared 100 instances of QKP with parameters $N=8, 16, 32, 64$ and $\Delta=0.2, 0.6, 1.0$. 
We performed our algorithm along with the comparison methods. 
The results of these comparative experiments are presented in the next chapter.

\subsection{results}
In this chapter, we assess the performance of the introduced algorithm using QKP. 
We solve QKP using the following four methods and compare their accuracy:

\begin{itemize}
  \item {\bf Ohzeki method with MCMC (OM(MCMC))} : Using Markov-chain Monte-Carlo (MCMC) method for sampling the expectation values.
  \item {\bf Ohzeki method with SQA (OM(SQA))} : Using simulated quantum annealing (SQA) for sampling the expectation values.
  \item {\bf Naive subgradient method} : A method that solves the Lagrangian relaxation problem \eqref{new objective function for inequality constraints} exactly without sampling via MCMC or SQA.
  \item {\bf Greedy method for QKP} : A heuristic approach specific to QKP.
\end{itemize}
For OM(MCMC), we use MCMC to sample the expectation values; for OM(SQA), we use SQA. We use OpenJij as the sampler for MCMC and SQA. For MCMC, the number of samples is set to $S=1000$, and for SQA, we set the Trotter number to $2$ and the number of samples to $S=500$. The inverse temperature is fixed at $\beta=0.1$ for both MCMC and SQA.

The naive subgradient method solves the Lagrangian relaxation problem \eqref{new objective function for inequality constraints} exactly rather than sampling using MCMC or SQA. 
This method corresponds to the simple combination of Lagrangian relaxation and subgradient methods. 
In this case, instead of calculating the expectation values by sampling, a single exact solution ${\bm x}^{(t)}_{\rm exact}$ is used. Therefore, in Eqs \eqref{sub-gradient update rule} and \eqref{sub-gradient eta update rule}, $\left\langle f_0({\bm x}) \right\rangle_{Q({\bm  \mu^{(t)}})}$ and $\left\langle F_k({\bm x}) \right\rangle_{Q({\bm  \mu^{(t)}})}$ are replaced by $f_0({\bm x}^{(t)}_{\rm exact})$ and $F_k({\bm x}^{(t)}_{\rm exact})$, respectively. The exact solution ${\bm x}^{(t)}_{\rm exact}$ is obtained using the Gurobi Optimizer (version 10.0.0).

The greedy method for QKP, used in the literature\cite{BILLIONNET1996310}, is the same as the method we use to calculate the upper bound $f_0^{\rm UB}({\bm x})$ for updating the step size $\eta^{(t)}$ in Eq. \eqref{sub-gradient eta update rule}.

We use the relative error $|f_0({\bm x}) - f_0({\bm x}^{\ast})| / |f_0({\bm x}^{\ast})|$ as the accuracy metric, where $f_0({\bm x}^{\ast})$ is the the objective function obtained by solving the QKP exactly using Eq. \eqref{QKP}. 
The exact solution $\bm x^{\ast}$ is obtained using the Gurobi Optimizer. 
Here $f_0({\bm x})$ is the objective function obtained by each method described above.

\begin{figure}
    \centering
    \includegraphics[width=1.0\linewidth]{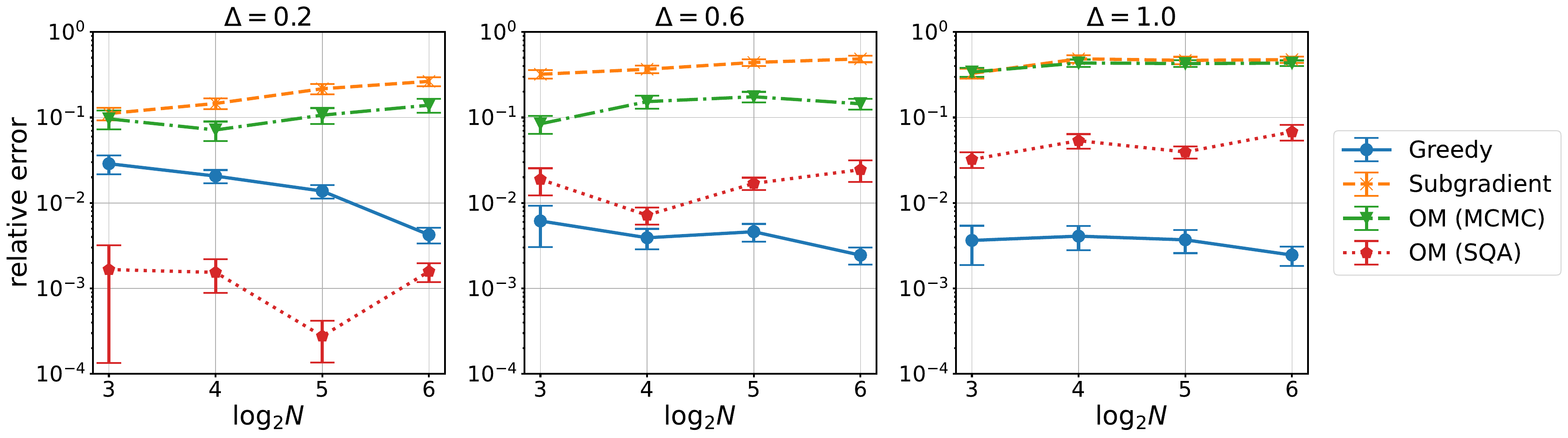}
    \caption{$N$-dependence of the relative error of each method at $\Delta=0.2, 0.6, 1.0$. The vertical axis represents the relative error. The horizontal axis denotes the problem size $N$.  }
    \label{fig:relative_error_N_dependence}
\end{figure}

Figure \ref{fig:relative_error_N_dependence} shows the $N-$dependence of the relative error for each method.  Each plot represents the average relative error over 100 instances, with error bars indicating the standard error.

The results show that when $\Delta=0.2$, SQA yields the best performance overall. As $\Delta$ increases, MCMC's accuracy degrades, eventually becoming nearly identical to the naive subgradient method at $\Delta=1.0$. SQA also declines in accuracy as $\Delta$ approaches 1.0, performing worse than the greedy method at $\Delta=0.6$ and $\Delta=1.0$. 

\begin{figure}
    \centering
    \includegraphics[width=1.0\linewidth]{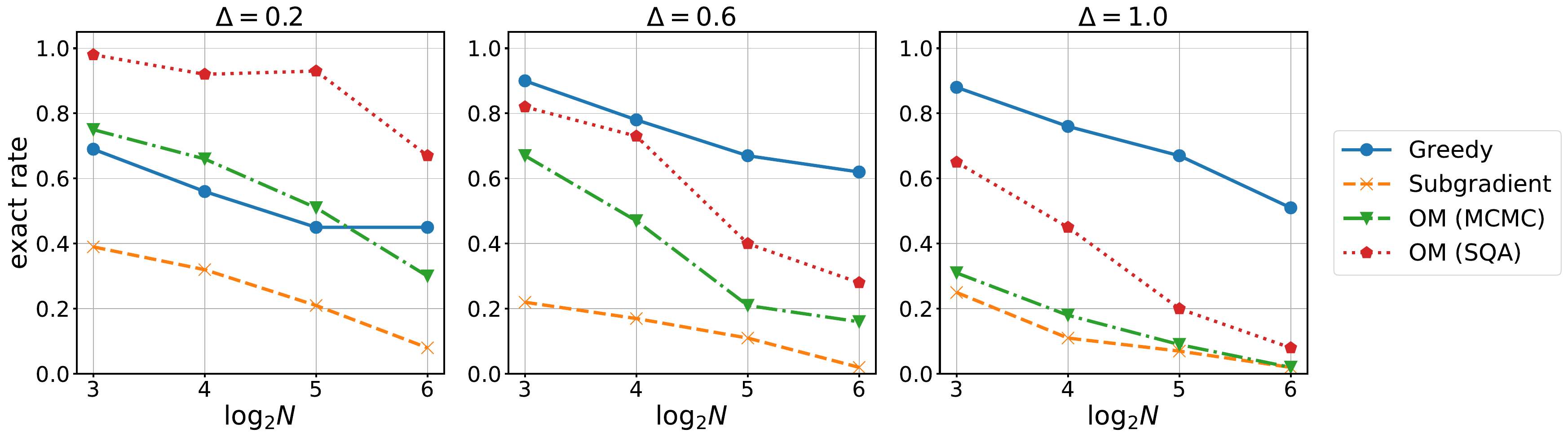}
    \caption{$N$-dependence of the exact rate (rate of reaching exact solutions) obtained by each method at $\Delta=0.2, 0.6, 1.0$. The vertical axis represents the exact rate. The horizontal axis denotes the problem size $N$.  }
    \label{fig:exact_rate_N_dependence}
\end{figure}

Figure \ref{fig:exact_rate_N_dependence} shows the exact rate, the proportion of instances among 100 instances where the exact solution was obtained. 
At $\Delta=0.2$, SQA achieves the highest exact rate. However, the exact rate decreases as $N$ increases, and for both MCMC and SQA, it also declines as $\Delta$ grows larger. When $\Delta$ reaches 0.6 and 1.0, SQA's exact rate falls below that of the greedy method.

\begin{figure}
    \centering
    \includegraphics[width=1.0\linewidth]{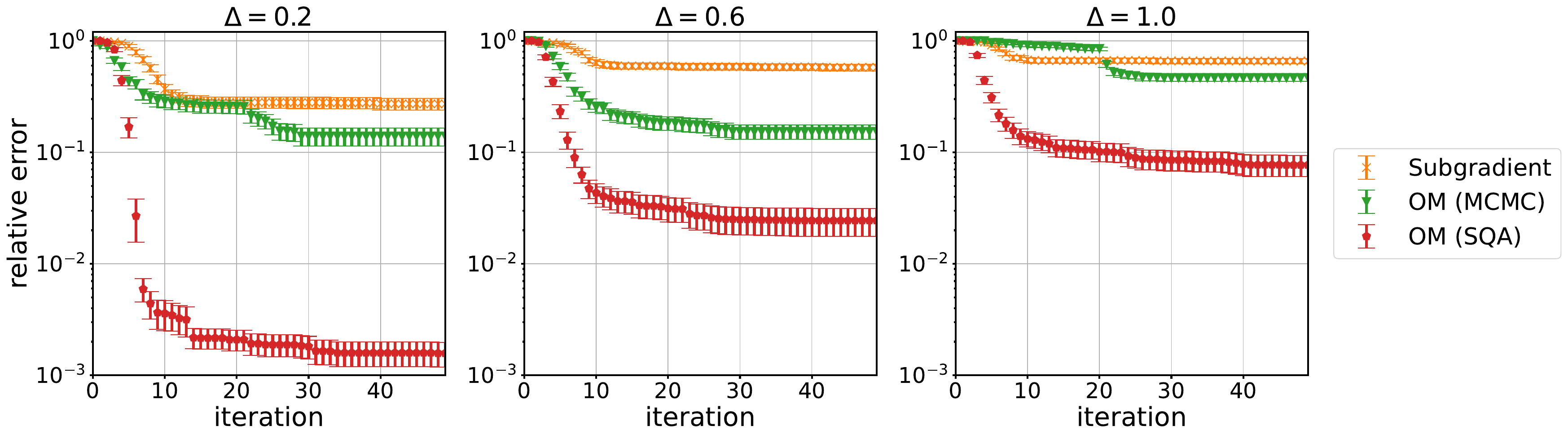}
    \caption{Relative error of each method at each iteration in QKP ($N=64$) with $\Delta=0.2, 0.6, 1.0$.The vertical axis represents the relative error. The horizontal axis denotes the iteration.  }
    \label{fig:iter_N_dependence}
\end{figure}

Figure \ref{fig:iter_N_dependence} illustrates the average relative error over iterations for OM(MCMC), OM(SQA), and the naive subgradient method. 
These results correspond to QKP instances with $N=64$ under different densities $\Delta=0.2, 0.6, 1.0$. 
For each instance, the relative error is set to 1 if a feasible solution is not found, and the relative error of the best solution is used if multiple feasible solutions are found. The results indicate that as the number of iterations increases, the relative error for the naive subgradient method diverges from that of OM(MCMC) and OM(SQA). 

\section{Discussion}
We derived the update rule \eqref{sub-gradient update rule} for the Ohzeki method applied to inequality-constrained optimization problems and confirmed that, except for the range of the artificial variable $\bm \mu$, it matches the update rule for the equality constraint case \eqref{update rule for equality constraint}. We investigated the performance of the Ohzeki method using the Markov-chain Monte-Carlo (MCMC) method and simulated quantum annealing (SQA) as samplers through experiments with the quadratic knapsack problem (QKP). 

The numerical experiments show that OM(MCMC) and OM(SQA) tend to be more accurate than the naive subgradient method, particularly when $\Delta$ is small. The difference between the Ohzeki method and the naive subgradient method lies in whether the parameter updates are based on expectation values obtained by sampling from the Lagrangian relaxation problem \eqref{new objective function for inequality constraints} or from a single exact solution. These results demonstrate the effectiveness of sampling in methods that combine Lagrangian relaxation with the subgradient method.

Under the condition $\beta=0.1$, OM(SQA) also achieves higher accuracy than OM(MCMC), presumably because the Trotter interactions of SQA promote sampling from lower-energy states more effectively than MCMC. This observation highlights a potential advantage of using QA as the sampler in the Ohzeki method. However, as $\Delta$ increases, both OM(SQA) and OM(MCMC) exhibit lower accuracy, and even OM(SQA) is outperformed by the greedy method at $\Delta=0.6$ and $\Delta=1.0$. This result indicates that sampling from the Lagrangian relaxation problem does not necessarily converge to better solutions. The QKP is characterized by a tendency for items with higher efficiency (the ratio of value to weight) to contribute to better solutions. The greedy method, which selects items based on this efficiency, is therefore well-suited to the structure of the QKP. In particular, when there is sufficient capacity, it becomes possible to select a larger number of high-efficiency items, leading to improved solutions\cite{10.1145/1068009.1068111, ohno_toward_2024}.

A related method for solving inequality-constrained optimization problems using quantum annealers is the augmented Lagrangian method\cite{Yonaga2020, djidjev_quantum_2023, cellini_qal-bp_2024}, which represents constraints using both linear and quadratic terms like $-\sum_{k=1}^{K}\alpha_{k}(F_k({\bm x})-C_k)+\frac{\beta}{2}\sum_{k=1}^{K}(F_k({\bm x})-C_k)^2$. 
The quadratic term enforces the constraint more strictly, making the solution more likely to satisfy $F_k({\bm x})-C_k=0$ than the Lagrangian relaxation method alone. 
This may lead to better convergence of the objective function $f_0({\bm x})$.

However, the augmented Lagrangian method can only handle constraints linear in the variables, such as $F_k({\bm x}) = \sum_{i=1}^{N} a_{k i} x_i$. 
In contrast, as discussed in the Sect. 3, the Ohzeki method can handle constraints with quadratic terms, such as $F_k({\bm x}) = \sum_{i=1}^{N} \sum_{j=1}^{N} A_{k i j} x_i x_j$. 
Another method proposed by Hirama handles problems with quadratic constraint terms\cite{hirama_efficient_2023}. 
It applies Dantzig-Wolfe decomposition to inequality-constrained optimization problems, reducing the problem to a QUBO that can be solved using quantum annealers. 
This QUBO corresponds to the one solved by the Ohzeki method \eqref{new objective function for inequality constraints}. 
However, in Hirama's method, QA is used as a part of the column generation process in a mathematical optimization algorithm rather than as the final solver. 
Therefore, a comparative study using the Ohzeki method could be considered for future work.

In addition, the objective function derived by the Ohzeki method for solving problems using quantum annealers \eqref{new objective function for inequality constraints} corresponds to the Lagrangian relaxation problem in mathematical optimization. 
When recalling that ${\bm \mu}$ in Eq. \eqref{new objective function for inequality constraints} is a variable in the imaginary axis, the corresponding optimization problem is known as the Lagrangian dual problem\cite{geoffrion_duality_1971}. 
In addition, in mathematical optimization, methods that combine Lagrangian relaxation with subgradient methods are often used to solve inequality-constrained optimization problems, and the update rule \eqref{sub-gradient update rule} aligns with these methods. 
These derived results suggest a connection between statistical physics and mathematical optimization. Exploring other methods in mathematical optimization with statistical physics could be an interesting avenue for future research.

{\it Acknowledgement.}
This study was financially supported by programs for bridging the gap between R\&D and IDeal society (Society 5.0) and Generating Economic and social value (BRIDGE) and Cross-ministerial Strategic Innovation Promotion Program (SIP) from the Cabinet Office.

\section*{Appendix A: Derivation of the Update Rule}

This appendix derives the update rule \eqref{sub-gradient update rule} for the Ohzeki method applied to inequality-constrained optimization problems. First, note that the saddle-point equation \eqref{saddle point inequality1} is satisfied when the constraint $k$ is not explicitly needed - that is, when $\langle F_k({\bm x}) \rangle_{Q({\bm \mu})} \le C_k$ holds for any distribution $Q({\bm \mu})$. In this case, Eq. \eqref{saddle point inequality1} implies $\mu_k=0$. Therefore, the penalty term $\mu_k F_k({\bm x})$ in the Lagrangian relaxation problem \eqref{new objective function for inequality constraints} vanishes.

Next, Eq. \eqref{saddle point inequality2} is satisfied when a certain distribution $Q({\bm \mu})$ with $\mu_k \ne 0$ still ensures $\langle F_k({\bm x}) \rangle_{Q({\bm \mu})}\le C_k$. In Eq. \eqref{saddle point inequality2}, $\xi_k\ge 0$ can be viewed as a slack variable that represents how much $\langle F_k({\bm x}) \rangle_{Q({\bm \mu})}$ is below $C_k$. Consequently, the inequality constraint $\langle F_k({\bm x}) \rangle_{Q({\bm \mu})}\le C_k$ is recast as the equality $\langle F_k({\bm x}) \rangle_{Q({\bm \mu})} + \xi_k=C_k$.

If $\langle F_k({\bm x}) \rangle_{Q({\bm \mu})} > C_k$, the saddle-point equation \eqref{saddle point inequality2} is not satisfied. In that situation, the slack variable must vanish, $\xi_k=0$, and the distribution $Q({\bm \mu})$ must be adjusted to reduce $\langle F_k({\bm x}) \rangle_{Q({\bm \mu})}$. From Eq. \eqref{saddle point inequality2}, the gradient for steepest-ascent is 
\begin{equation}
    \frac{\partial}{\partial \mu_k}H(\mu_k, \xi_k)=\langle F_k({\bm x}) \rangle_{Q({\bm \mu})}-C_k.
\end{equation}

We also impose $\mu_k\ge 0$. Allowing $\mu_k< 0$ would undermine the role of $\sum_k \mu_k (F_k({\bm x}) - C_k)$ as a proper penalty term, because $\mu_k < 0$ could cause the Hamiltonian to decrease when $F_k({\bm x})$ increases. Consequently, $\mu_k\ge 0$ is necessary so that the solution moves toward satisfying the constraints. Combining these observations yields the update rule \eqref{sub-gradient update rule}. This rule is equivalent to the subgradient ascent method applied to the Lagrangian relaxation problem with inequality constraints.

Hence, if a distribution $Q({\bm \mu})$ satisfies $\langle F_k({\bm x}) \rangle_{Q({\bm \mu})} \le C_k$, then we must have $\mu_k\ge0, \xi_k\ge0$, and $\mu_k(\xi_k - C_k + \langle F_k({\bm x}) \rangle_{Q({\bm \mu})})=0$. Interpreting $\mu_k$ as the Lagrange multiplier and $\xi_k$ as the slack variable shows that these conditions fulfill the Karush-Kuhn-Tucker (KKT) conditions, especially the complementary slackness condition.

In the thermodynamic limit $\beta\rightarrow+\infty$, the main contribution in the integral of Eq. \eqref{partition function after variable transformation} arises near the saddle point (stationary point). Since the transformation $z_k \rightarrow -i\beta \mu_k$ moves $\bm \mu$ onto the imaginary axis, the maximization with respect to $\bm \mu$ in Eq. \eqref{effective hamiltonian for inequality constraints} corresponds to the Lagrangian dual problem. The saddle-point condition is enforced by iterating the update rule \eqref{sub-gradient update rule} to drive $Q({\bm \mu})$ to a distribution that ensures the integral converges to a finite real value.

\section*{Appendix B: Comparing the Number of Quadratic Terms}
\begin{figure}
    \centering
    \includegraphics[width=1.0\linewidth]{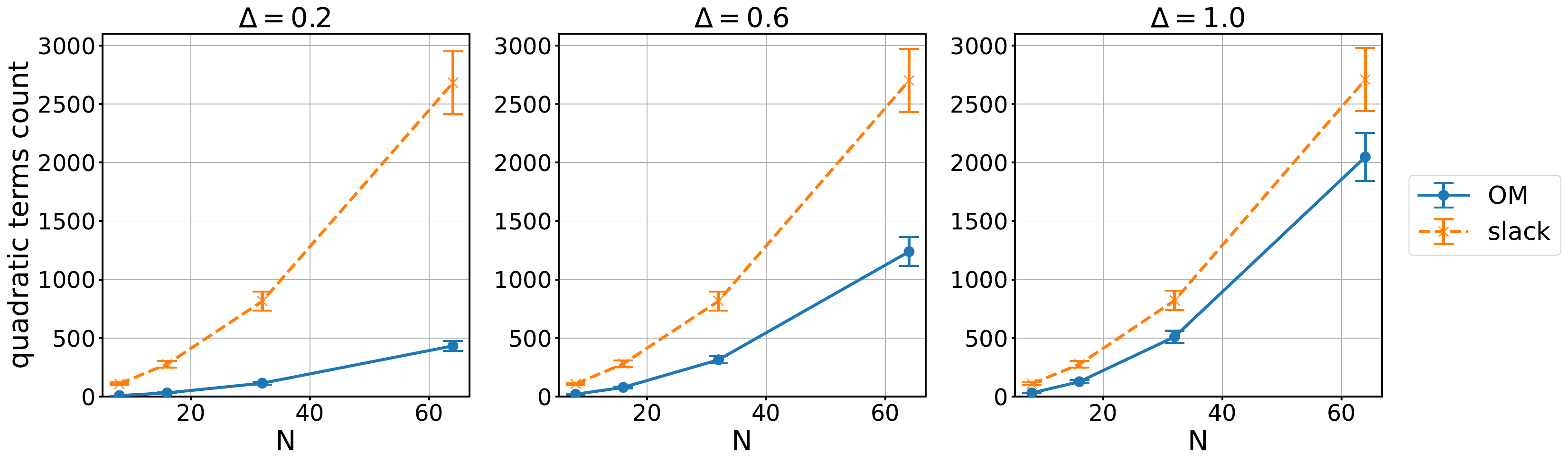}
    \caption{$N$-dependence of the number of quadratic terms of each method at $\Delta=0.2, 0.6, 1.0$. The vertical axis represents the number of quadratic terms. The horizontal axis denotes the problem size $N$.  }
    \label{fig:quadratic_terms_count}
\end{figure}

Here, we compare the number of quadratic terms in the QUBO derived from the Ohzeki method in Eq. \eqref{new objective function for inequality constraints} to that in the conventional slack-variable approach for handling inequality constraints. For this comparison, we use the quadratic knapsack problem (QKP) in Eq. \eqref{QKP}. We generate 100 random instances for each $N=8, 16, 32, 64$ and densities $\Delta = 0.2, 0.6, 1.0$, following the procedure in Sect. 4. Figure \ref{fig:quadratic_terms_count} shows how the number of quadratic terms depends on $N$ for each $\Delta$.

From Fig. \ref{fig:quadratic_terms_count}, both the Ohzeki method (labeled "OM") and the slack-variable approach have more quadratic terms as $N$ grows. However, the growth rate in the Ohzeki method is smaller than that in the slack-variable approach. This difference arises because the Ohzeki method converts constraints into linear terms, whereas the slack-variable approach introduces extra binary variables and couples them quadratically to all existing variables. Consequently, the slack-variable method produces a fully connected penalty structure regardless of $\Delta$. By contrast, in the Ohzeki method, only the original objective part $f_0({\bm x})$ (which can vary in density depending on $\Delta$) contributes to additional connections.

In summary, the Ohzeki method reduces the number of quadratic terms compared with the conventional slack-variable approach. This reduction simplifies embedding into quantum annealers with sparse connections and can help preserve solution quality and increase the feasible problem size.

* taisei.takabayashi.s1@dc.tohoku.ac.jp

\bibliographystyle{jpsj}
\bibliography{main}

\begin{thebibliography}{10}

\bibitem{kadowaki1998quantum}
T.~Kadowaki and H.~Nishimori: Physical Review E {\bfseries 58} (1998) 5355.

\bibitem{neukart2017traffic}
F.~Neukart, G.~Compostella, C.~Seidel, D.~Von~Dollen, S.~Yarkoni, and B.~Parney: Front. ICT {\bfseries 4} (2017) 29.

\bibitem{inoue2021traffic}
D.~Inoue, A.~Okada, T.~Matsumori, K.~Aihara, and H.~Yoshida: Scientific reports {\bfseries 11} (2021) 1.

\bibitem{shikanai2023trafficsignaloptimizationusing}
R.~Shikanai, M.~Ohzeki, and K.~Tanaka.
\newblock Traffic signal optimization using quantum annealing on real map, 2023.

\bibitem{rosenberg2016solving}
G.~Rosenberg, P.~Haghnegahdar, P.~Goddard, P.~Carr, K.~Wu, and M.~L. De~Prado: IEEE J. Sel. Top. Signal Process. {\bfseries 10} (2016) 1053.

\bibitem{venturelli2019reverse}
D.~Venturelli and A.~Kondratyev: Quantum Machine Intelligence {\bfseries 1} (2019) 17.

\bibitem{Ishikawa_2023}
Y.~Ishikawa, T.~Yoshihara, K.~Okamura, and M.~Ohzeki: Frontiers in Computer Science {\bfseries 5} (2023).

\bibitem{ohzeki2019control}
M.~Ohzeki, A.~Miki, M.~J. Miyama, and M.~Terabe: Frontiers in Computer Science {\bfseries 1} (2019) 9.

\bibitem{Haba2022}
R.~Haba, M.~Ohzeki, and K.~Tanaka: Scientific Reports {\bfseries 12} (2022) 17753.

\bibitem{Tanaka_2023}
T.~Tanaka, M.~Sako, M.~Chiba, C.~Lee, H.~Cha, and M.~Ohzeki: Journal of the Physical Society of Japan {\bfseries 92} (2023).

\bibitem{Doi_2023}
M.~Doi, Y.~Nakao, T.~Tanaka, M.~Sako, and M.~Ohzeki: Frontiers in Computer Science {\bfseries 5} (2023).

\bibitem{yonaga_quantum_2022}
K.~Yonaga, M.~Miyama, M.~Ohzeki, K.~Hirano, H.~Kobayashi, and T.~Kurokawa: ISIJ International {\bfseries 62} (2022) 1874.

\bibitem{ide2020maximum}
N.~Ide, T.~Asayama, H.~Ueno, and M.~Ohzeki: 2020 International Symposium on Information Theory and Its Applications (ISITA), 2020, pp. 91--95.

\bibitem{Arai2021code}
S.~Arai, M.~Ohzeki, and K.~Tanaka: Phys. Rev. Research {\bfseries 3} (2021) 033006.

\bibitem{amin2018}
M.~H. Amin, E.~Andriyash, J.~Rolfe, B.~Kulchytskyy, and R.~Melko: Physical Review X {\bfseries 8} (2018).

\bibitem{o2018nonnegative}
D.~O’Malley, V.~V. Vesselinov, B.~S. Alexandrov, and L.~B. Alexandrov: PloS one {\bfseries 13} (2018) e0206653.

\bibitem{sato_assessment_2021}
T.~Sato, M.~Ohzeki, and K.~Tanaka: Scientific Reports {\bfseries 11} (2021) 13523.
\newblock Publisher: Nature Publishing Group.

\bibitem{Urushibata2022}
M.~Urushibata, M.~Ohzeki, and K.~Tanaka: Journal of the Physical Society of Japan {\bfseries 91} (2022) 074008.

\bibitem{hasegawa2023}
Y.~Hasegawa, H.~Oshiyama, and M.~Ohzeki.
\newblock Kernel Learning by quantum annealer, 2023.

\bibitem{goto2023onlinecalibrationschemetraining}
T.~Goto and M.~Ohzeki.
\newblock Online calibration scheme for training restricted Boltzmann machines with quantum annealing, 2023.

\bibitem{Koshikawa2021}
A.~S. Koshikawa, M.~Ohzeki, T.~Kadowaki, and K.~Tanaka: J. Phys. Soc. Jpn. {\bfseries 90} (2021) 064001.

\bibitem{Oshiyama2022}
H.~Oshiyama and M.~Ohzeki: Sci. Rep. {\bfseries 12} (2022) 2146.

\bibitem{Yamamoto2020}
M.~Yamamoto, M.~Ohzeki, and K.~Tanaka: J. Phys. Soc. Jpn. {\bfseries 89} (2020) 025002.

\bibitem{Maruyama2021}
N.~Maruyama, M.~Ohzeki, and K.~Tanaka:   (2021).

\bibitem{glover_quantum_2022}
F.~Glover, G.~Kochenberger, R.~Hennig, and Y.~Du: Annals of Operations Research {\bfseries 314} (2022) 141.

\bibitem{Ohzeki2020}
M.~Ohzeki: Scientific Reports {\bfseries 10} (2020) 3126.

\bibitem{Hubbard1959}
J.~Hubbard: Phys. Rev. Lett. {\bfseries 3} (1959) 77.

\bibitem{Stratonovich1957}
R.~L. {Stratonovich}: Soviet Physics Doklady {\bfseries 2} (1957) 416.

\bibitem{geoffrion_lagrangean_1974}
A.~M. Geoffrion, Lagrangean relaxation for integer programming, In M.~L. Balinski (ed), {\em Approaches to {Integer} {Programming}}, pp. 82--114. Springer Berlin Heidelberg, Berlin, Heidelberg, 1974.

\bibitem{Kirkpatrick1983}
S.~Kirkpatrick, C.~D. Gelatt, and M.~P. Vecchi: Science {\bfseries 220} (1983) 671.

\bibitem{suzuki_monte_1977}
M.~Suzuki, S.~Miyashita, and A.~Kuroda: Progress of Theoretical Physics {\bfseries 58} (1977) 1377.

\bibitem{BILLIONNET1996310}
A.~Billionnet and F.~Calmels: European Journal of Operational Research {\bfseries 92} (1996) 310.

\bibitem{fisher_lagrangian_1981}
M.~L. Fisher: Management Science {\bfseries 27} (1981) 1.
\newblock Publisher: INFORMS.

\bibitem{Shunji_Umetani2007KJ00004805415}
S.~Umetani and M.~Yagiura: Journal of the Operations Research Society of Japan {\bfseries 50} (2007) 350.

\bibitem{gallo_quadratic_1980}
G.~Gallo, P.~L. Hammer, and B.~Simeone, Quadratic knapsack problems, In M.~W. Padberg (ed), {\em Combinatorial {Optimization}}, pp. 132--149. Springer, Berlin, Heidelberg, 1980.

\bibitem{10.1145/1068009.1068111}
B.~A. Julstrom: Proceedings of the 7th Annual Conference on Genetic and Evolutionary Computation, GECCO '05, 2005, p. 607–614.

\bibitem{ohno_toward_2024}
K.~Ohno, T.~Shirai, and N.~Togawa: IEEE Access {\bfseries 12} (2024) 97678.
\newblock Conference Name: IEEE Access.

\bibitem{Yonaga2020}
K.~Yonaga, M.~J. Miyama, and M.~Ohzeki: arXiv:2012.06119 .

\bibitem{djidjev_quantum_2023}
H.~N. Djidjev: Advanced Quantum Technologies {\bfseries 6} (2023) 2300104.
\newblock \_eprint: https://onlinelibrary.wiley.com/doi/pdf/10.1002/qute.202300104.

\bibitem{cellini_qal-bp_2024}
L.~Cellini, A.~Macaluso, and M.~Lombardi: Scientific Reports {\bfseries 14} (2024) 5142.
\newblock Publisher: Nature Publishing Group.

\bibitem{hirama_efficient_2023}
S.~Hirama and M.~Ohzeki: Journal of the Physical Society of Japan {\bfseries 92} (2023) 113002.
\newblock Publisher: The Physical Society of Japan.

\bibitem{geoffrion_duality_1971}
A.~M. Geoffrion: SIAM Review {\bfseries 13} (1971) 1.
\newblock \_eprint: https://doi.org/10.1137/1013001.

\end{thebibliography}
\end{document}